\documentclass[12pt]{article}
\usepackage{amsmath, amssymb, amsthm}
\usepackage{mathrsfs, bm}
\usepackage{graphicx}
\usepackage{subcaption, float}
\usepackage{caption}
\usepackage{cleveref, cite, indentfirst, comment, enumerate}
\usepackage[a4paper, left=3cm, right=3cm, top=4.1cm, bottom=4.1cm]{geometry}
\usepackage{titlesec}

\newtheoremstyle{myplain}{\topsep}{\topsep}{}{}{\bfseries}{}{0.1em}
{\thmname{#1}\thmnumber{ #2}\thmnote{ #3} }   

\theoremstyle{myplain}
\newtheorem{theorem}{Theorem}[section]
\newtheorem{lemma}[theorem]{Lemma}
\newtheorem{corollary}[theorem]{Corollary}

\newtheorem{property}{Property}
\numberwithin{equation}{section}

\titleformat{\section}
{\normalfont\Large\bfseries}
{\thesection.}
{0.5em}
{}

\title{On the Complexity of Extending Prescribed  Linear Forests to Hamiltonian Cycles in Hypercubes\thanks{This research was partially supported by the National Natural Science Foundation of China (No. 12261071)}}

\author{Huazhong L\"{u}$^{1}$\thanks{Corresponding author.} and Tingzeng Wu$^{2}$\\
{\small $^{1}$School of Mathematical Sciences, } \\
{\small University of Electronic Science and Technology of China,} \\
{\small Chengdu, Sichuan 611731, P.R. China}\\
{\small E-mail: lvhz@uestc.edu.cn}\\
{\small $^{2}$School of Mathematics and Statistics, Qinghai Nationalities University, }\\
{\small Xining, Qinghai 810007, P.R. China} \\
{\small E-mail: mathtzwu@163.com}\\}
\date{}

\begin{document}
\maketitle
\begin{abstract}

Let $Q_n$ be the $n$-dimensional hypercube, and let $P\subseteq E(Q_n)$ be a set of prescribed edges. It is known that, for $n\ge 2$, every such set $P$ with $|P|\leq 2n-3$ whose edges form a linear forest is contained in a Hamiltonian cycle of $Q_n$. This bound is sharp for general linear forests. The computational complexity of deciding whether a prescribed edge set of a hypercube can be extended to a Hamiltonian cycle was posed as an open question by Dvořák [SIAM J. Discrete Math. 19 (2005) 135--144]. We prove that the problem is NP-complete even when the prescribed edges form a spanning linear forest with no isolated vertices. We give a polynomial-time reduction from a restricted 3-SAT problem, using local spanning path-forest gadgets for variables and clauses that are uniformly embedded in the hypercube. The same reduction therefore establishes NP-hardness of two versions of Hamiltonian-cycle extension problems for general prescribed edge sets.

\vskip 0.1 in
		

\end{abstract}

\section{Introduction}\label{sec:introduction}

The $n$-dimensional hypercube $Q_n$ is the graph on the binary words of length $n$, in which two vertices are adjacent if they differ in exactly one coordinate. Its Hamiltonian cycles, equivalently cyclic binary Gray codes, have been studied extensively in combinatorics and interconnection-network theory \cite{Savage1997,ChenShin1987}. Although $Q_n$ has a Hamiltonian cycle for every $n\ge 2$, additional requirements on the edges of such a cycle lead to substantially different questions. Given a set $P\subseteq E(Q_n)$, the problem is to find a Hamiltonian cycle of $Q_n$ containing every edge of $P$. If such a cycle exists, then the graph formed by $P$ must be a linear forest, unless $P$ itself is a Hamiltonian cycle. It is therefore natural to ask whether this necessary condition is sufficient for a Hamiltonian extension and, if not, what additional restrictions are needed.

Caha and Koubek~\cite{CahaKoubek2006} proved that every set $P$ of at most $n-1$ edges of $Q_n$ whose edges form a collection of pairwise vertex-disjoint paths is contained in a Hamiltonian cycle of $Q_n$. Dvořák~\cite{Dvorak2005} established the bound $2n-3$: for $n\ge 2$ and $|P|\le 2n-3$, the set $P$ is contained in a Hamiltonian cycle of $Q_n$ if and only if it forms a linear forest. The bound is sharp for every $n\ge 3$, as there exists a linear forest in $Q_n$ with exactly $2n-2$ edges that is not contained in any Hamiltonian cycle.

Prescribed-edge extension to Hamiltonian path and cycle has also been studied in hypercubes and other interconnection networks. For hypercubes, Dvo\v{r}\'ak and Gregor \cite{DvorakGregor2007} characterized when a prescribed edge set $P$ with $|P|\le 2n-4$ can be contained in a Hamiltonian path of $Q_n$ with specified end vertices. For $k$-ary $n$-cubes, Stewart~\cite{Stewart2011} proved that every prescribed linear forest with at most $2n-1$ edges extends to a Hamiltonian cycle when $k\ge 4$. Yang and Zhang~\cite{YangZhang2022TPDS} studied the extension of prescribed linear forests to Hamiltonian paths with specified end vertices in $k$-ary $n$-cubes with odd $k$, in the presence of faulty edges. For balanced hypercubes, L\"{u} and Wang \cite{LuWang2019Balanced} proved that every prescribed linear forest with at most $n$ edges extends to a Hamiltonian cycle, and Cheng~\cite{Cheng2019BalancedPrescribed} subsequently improved this bound to $2n-1$.

A particularly important special case arises when the prescribed edge set is a matching. In this setting, the classical Ruskey-Savage conjecture \cite{RuskeySavage1993} asks whether every matching of a hypercube can be extended to a Hamiltonian cycle. Although the conjecture remains open, Hamiltonian-cycle extension has been established for several restricted classes of matchings, including matchings of size up to specified bounds and matchings whose edges lie in at most five dimensions. Fink~\cite{Fink2007} proved that every perfect matching of $Q_n$ can be extended to a Hamiltonian cycle, thereby settling Kreweras' conjecture. He also pointed out that every matching of $Q_n$ can be extended to a Hamiltonian cycle when $n\in\{2,3,4\}$. Wang and Zhao~\cite{FWWZ} subsequently established the conjecture for $Q_5$.

For general matchings, several works have established size bounds that guarantee a Hamiltonian extension. Wang and Zhang~\cite{WangZhang2014} proved that every matching of size at most $2n-1$ has a Hamiltonian extension. They later improved this bound to $3n-10$ in a subsequent paper~\cite{WangZhang2016}. Dvo\v{r}\'ak and Fink~\cite{TDJF} established the quadratic bound $\frac{n^{2}}{16}+\frac{n}{4}$, a bound that grows quadratically with the dimension $n$. More recently, Ali et al. \cite{AliAliYang2026} proved that, for $n\ge 6$, every matching $M$ in $Q_n$ with $|M|\le 4n-21$ extends to a Hamiltonian cycle. In a different direction, Wang and Wang~\cite{SWFW} proved that a matching $M$ of $Q_n$ extends to a Hamiltonian cycle if its edges lie in at most five dimensions and $|M|<10\times 2^{n-5}$. Fink and Hotmar~\cite{FinkHotmar2026} subsequently removed the size condition, establishing Hamiltonian extension for every matching whose edges lie in at most five dimensions. A complementary sufficient condition for Hamiltonian extension, based on local bounds on the number of matching edges in subcubes, was established in \cite{Zuolv2026}. Fink and M\"utze~\cite{FinkMutze2025} also proved that every matching of $Q_n$, $n\ge2$, can be extended to a cycle that visits at least a $2/3$-fraction of all vertices. Although these works establish bounds on the size of matchings that are guaranteed to extend to Hamiltonian cycles, as well as structural conditions under which such extensions exist, they do not by themselves provide a general criterion for extending arbitrary prescribed edge sets to Hamiltonian cycles.

A related problem asks whether $Q_n-F$ is Hamiltonian for an arbitrary set $F\subseteq E(Q_n)$ of faulty edges. Chan and Lee~\cite{ChanLee1991} proved that this problem is NP-complete. In contrast, the prescribed-edge problem requires a Hamiltonian cycle containing every edge of a given set $P$, rather than avoiding a set of forbidden edges.

Dvo\v{r}\'ak \cite{Dvorak2005} raised the question of whether Hamiltonian-cycle extension through prescribed edges in hypercubes is also NP-complete.
Stewart \cite{Stewart2011} later observed that the unrestricted prescribed-edge problem for hypercubes and $k$-ary $n$-cubes had not yet been investigated from a complexity-theoretic perspective. To the best of our knowledge, the complexity of deciding whether an arbitrary prescribed edge set in a hypercube can be extended to a Hamiltonian cycle remains unresolved.

We study the following Hamiltonian-cycle extension problem in hypercubes.

\smallskip
\noindent\textsc{Spanning Linear-Forest Hamiltonian-Cycle Extension}

\smallskip
\noindent\textit{Instance:} An integer $n\ge 2$, encoded in binary, and a set $P\subseteq E(Q_n)$ represented by an explicit list of edges, each specified by its two $n$-bit endpoints, such that the graph $(V(Q_n),P)$ is a spanning linear forest with no isolated vertices. In this representation, $Q_n$ is specified implicitly by its dimension $n$, whereas $P$ is specified explicitly by its edge list. 

\smallskip
\noindent\textit{Question:} Does $Q_n$ contain a Hamiltonian cycle $C$ such that $P\subseteq E(C)$?

\medskip
Our main theorem is the following.

\begin{theorem}\label{thm:main}
The \textsc{spanning linear-forest hamiltonian-cycle extension} problem is NP-complete.
\end{theorem}

Furthermore, the same reduction establishes
NP-hardness for the Hamiltonian-cycle extension
problem when the prescribed edge set is either
an arbitrary linear forest or an unrestricted
edge set (see Corollary~\ref{cor:broader}).

In our input representation, the hypercube $Q_n$ is specified implicitly by its dimension $n$, whereas the prescribed edge set $P$ is given explicitly. This distinction is important for establishing membership in $\mathrm{NP}$. For every instance of the restricted problem, the spanning and no-isolated-vertex conditions imply
$2|P|=\sum_{v\in V(Q_n)}\deg_P(v)\geq 2^n$. Since each edge of $P$ is represented by its two $n$-bit endpoints, the input length is $\Omega(n|P|)=\Omega(n2^n)$. On the other hand, a Hamiltonian cycle can be represented by listing its $2^n$ vertices, each as an $n$-bit word, requiring $O(n2^n)$ bits. Thus, an explicit Hamiltonian cycle provides a polynomial-size certificate for the restricted problem.

For general prescribed edge sets, however, $P$ may be sparse, so an explicit listing of a Hamiltonian cycle may have length superpolynomial in the input size. Consequently, the above certificate argument does not establish membership in $\mathrm{NP}$ for the unrestricted version.

The remainder of the paper is organized as follows. Section 2 sets up the reduction by specifying the source 3-SAT instances and describing the local hypercube constructions. Section 3 develops the uniform embedding of these constructions and verifies the required structural properties of the prescribed edge set. Section 4 establishes the soundness of the reduction. Section 5 completes the proof of Theorem~\ref{thm:main} and establishes two further NP-hardness results for general versions of the Hamiltonian-cycle extension problem. Section 6 concludes the paper. 

\section{Preliminaries and Source Problem}
\subsection{Notation, Input Representation, and Source Problem}

For a graph $G$, a path whose vertices appear in the order $v_0,v_1,\ldots,v_k$
is denoted by $\langle v_0,v_1,\ldots,v_k\rangle$. If $P=\langle v_0,\ldots,v_k\rangle$ and $Q=\langle w_0,\ldots,w_\ell\rangle$ with $v_k=w_0$, their concatenation is denoted by $\langle v_0,\ldots,v_k,w_1,\ldots,w_\ell\rangle$. A linear forest is a forest whose connected
components are paths. Equivalently, it is
an acyclic graph of maximum degree at most
two. Thus, the condition that
$(V(Q_n),P)$ is a spanning linear forest
with no isolated vertices is equivalent
to requiring that $P$ be acyclic and
$1\leq \deg_P(v)\leq 2$ for every
$v\in V(Q_n)$.

For two binary words $x,y\in\{0,1\}^{n}$, their Hamming distance is defined by $d_H(x,y)=|\{i\in[n]:x_i\neq y_i\}|$. The $n$-dimensional hypercube $Q_n$ is the graph with vertex set $V(Q_n)=\{0,1\}^{n}$ and edge set $E(Q_n)=\bigl\{\{x,y\}:x,y\in\{0,1\}^{n},\ d_H(x,y)=1\bigr\}$. For any integer $r\ge 1$ and any edge set $S\subseteq E(Q_r)$, we write $G_S=(V(Q_r),S)$ and denote the degree of a vertex $v$ in $G_S$ by $\deg_S(v)$.

The reduction starts from the following restricted
version of 3-SAT. A Boolean formula is
a conjunction of clauses, where each clause is a
disjunction of literals. In 3-SAT, every clause
contains exactly three literals. We use the
standard restriction in which the three literals
of each clause involve distinct variables. This
restriction remains NP-complete; see
Karp~\cite{Karp1972} and
Papadimitriou~\cite{Papadimitriou1994}.

Let $\Phi$ be an instance of this restricted version of 3-SAT with clauses $C_1,\ldots,C_m$. Without loss of generality, we discard any variables that do not appear in $\Phi$ and relabel the occurring variables as $x_1,\ldots,x_t$ in a fixed order, where $m\ge 1$ and $3\le t\le 3m$. For each clause $C_q$, write $C_q=\ell_{q,1}\vee\ell_{q,2}\vee\ell_{q,3}$, $q\in[m]$. Here $\ell_{q,k}$ denotes the $k$-th literal occurrence of clause $C_q$. For each pair $(q,k)\in[m]\times\{1,2,3\}$, let $i(q,k)\in[t]$ denote the index of the variable appearing in $\ell_{q,k}$, and let $b(q,k)\in\{0,1\}$ denote the truth value assigned to the variable $x_{i(q,k)}$ that makes $\ell_{q,k}$ true. The reduction constructs a hypercube instance from $\Phi$. The encoding of clauses, variables, and their connections is described in the following section.

\subsection{Path-Forest Expansion Operations}
 
The constructions below repeatedly enlarge a path forest by replacing an internal edge with a prescribed path in a larger subcube. We formalize this operation as a path-forest expansion.

For later use, we recall the binary reflected Gray path. For $d\ge1$ and $0\le k<2^d$, define
$g_k=k\oplus(k\gg1)$, where $\oplus$ denotes bitwise XOR and $k\gg1$ denotes shifting the binary representation of $k$ one position to the right. The sequence $\langle g_0,g_1,\ldots,g_{2^d-1}\rangle$ is the binary reflected Gray path of $Q_d$.

Throughout this subsection, we consider a spanning linear forest $F\subseteq E(Q_d)$ with no isolated vertices. An edge $e=uv\in F$ is called
an internal edge if $\deg_F(u)=\deg_F(v)=2$. A path-forest expansion at an internal edge $e=\{u,v\}$ is performed by introducing a new
half-cube adjacent to $Q_d$. Let $u'$ and $v'$ be the copies of $u$ and $v$ in this new
half-cube, and let $e'=\{u',v'\}$ be the corresponding parallel edge. The two edges $e$ and $e'$ lie in a 4-cycle. The expansion removes the edge $e$ and inserts the
path $\langle u,P(u',v'),v\rangle$, where $P(u',v')$ is the prescribed Hamiltonian
path in the newly introduced half-cube. 

We identify the vertices of $Q_d$ with their integer labels. Namely, a binary word $a_{d-1}\cdots a_1a_0\in\{0,1\}^d$, where $a_0$ is the least significant bit, is represented by the integer $\sum_{j=0}^{d-1}a_j2^j$. Let $ P_1,\ldots,P_s$
denote the path components of $G_F$, and
let $a_i$ and $b_i$ be the two endpoints
of $P_i$. We define $\operatorname{End}(F)=\{\{a_i,b_i\}:1\le i\le s\}$. Thus, $\operatorname{End}(F)$ records the unordered endpoint pairs of all path components of $F$. We also define the residual endpoint edge set as $R(F)=\{\{u,v\}\in E(Q_d):\deg_F(u)=\deg_F(v)=1,\ \{u,v\}\notin F\}$. Hence, $R(F)$ consists of cube edges that can potentially join two endpoints of the path forest without using an already prescribed edge.

The following properties follow immediately from
the definition of a path-forest expansion.

\begin{property}\label{prop:expansion}
Suppose that $e=\{u,v\}$ is an internal edge of $F$. Let $F'$ denote the spanning linear forest obtained by applying the path-forest expansion to $F$ at $e$. Then the following properties hold.

\begin{enumerate}
\item $(V(Q_{d+1}),F')$ is again a spanning
linear forest of $Q_{d+1}$.

\item Every path component of $F$ not containing
$e$ remains unchanged.

\item $\operatorname{End}(F')=\operatorname{End}(F)$.

\item $R(F')=R(F)$.
\end{enumerate}
\end{property}

\subsection{Path-Forest Endpoints and Completion Edges}

\begin{lemma}\label{lem:A1}
There exists a spanning linear forest $A_8\subseteq E(Q_8)$ consisting of two paths, with $\operatorname{End}(A_8)=\{\{14,38\},\{46,54\}\}$ and $R(A_8)=\{\{14,46\},\{38,46\},\{38,54\}\}$. Moreover, the edge $e=\{150,182\}\in A_8$ is an internal edge. After deleting $e$, the resulting forest satisfies $\operatorname{End}(A_8\setminus\{e\})=\{\{14,150\},\{38,182\},\{46,54\}\}$.
\end{lemma}

\begin{figure}[t]
    \centering
    \includegraphics[
        width=0.7\textwidth,
        keepaspectratio]{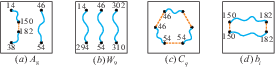}
    \caption{Local spanning path forests. (a) The two paths of $A_8$, with the specified edge $\{150,182\}$ marked; (b) The three paths of $W_9$, with endpoint pairs $\{14,294\}$, $\{46,54\}$, and $\{302,310\}$; (c) A three-cut clause forest, with its three cut edges shown dashed; (d) One possible pairing in a two-cut bridge forest, with ports $150$ and $182$ on each side.}
    \label{fig:A-8W-9}
\end{figure}

\begin{proof}
We construct $A_8$ from a two-path spanning forest
of $Q_4$ by four successive path-forest
expansions.

The initial forest $F_4$ consists of the two paths
$\langle 0,1,3,2,10,11,9,8,12,13,5\rangle$ and
$\langle 4,6,14,15,7\rangle$. Hence its endpoint pairs are $\operatorname{End}(F_4)=\{\{0,5\},\{4,7\}\}$. We apply path-forest expansions to the successive internal edges $\{1,3\}$, $\{1,17\}$, $\{1,33\}$ and $\{1,65\}$ appearing in the four expansion steps, when the current hypercube dimensions are $4,5,6,7$, respectively. At each step, the selected edge is replaced by the prescribed path in the newly introduced half-cube according to the definition of a path-forest expansion.

Let $F_8$ denote the resulting two-path spanning
forest in $Q_8$. By the structural properties of
path-forest expansions, the endpoint pairs are
preserved throughout the construction. Therefore,
when the vertices of the original $Q_4$ are viewed as the corresponding vertices of $Q_8$ with four leading zero coordinates, $\operatorname{End}(F_8)=\{\{0,5\},\{4,7\}\}$.

We now apply a cube automorphism only to obtain a
convenient labeling of this construction. Let $\pi$ be the coordinate permutation given by
\[
(\pi(0),\pi(1),\ldots,\pi(7))
=
(3,4,5,0,1,2,6,7),
\]
and for $x\in V(Q_8)$ define
\[
f(x)=14\oplus
\bigoplus_{j:\operatorname{bit}_j(x)=1}2^{\pi(j)},
\]
where $\operatorname{bit}_j(x)$ denotes the $j$-th bit of the binary representation of $x$.
Thus, $f$ consists of a permutation of the
coordinates followed by an XOR translation by the
vertex $14$. Hence, $f$ is an automorphism of
$Q_8$ and preserves adjacency.

A direct calculation from the definition of $f$
gives
\[
f(0)=14,\qquad f(5)=38,\qquad
f(4)=46,\qquad f(7)=54 .
\]
Therefore, $\operatorname{End}(F_8)$ is mapped
as $\{0,5\}\mapsto\{14,38\}$ and 
$\{4,7\}\mapsto\{46,54\}$. We define $A_8=f(F_8)$. Consequently, $\operatorname{End}(A_8)=\{\{14,38\},\{46,54\}\}$.

It remains to verify the existence and position of
the specified internal edge $e=\{150,182\}$.
Under the inverse of the above automorphism, $e$
is mapped to $\{131,135\}$.

The fourth path-forest expansion replaces the edge $\{1,65\}$ by a path entering the newly introduced half-cube $128+Q_7$. The inserted part is the Hamiltonian path obtained
by the XOR translation by the vertex $1$ of the
binary reflected Gray path defined above. Hence the vertices of the translated path are given by
$h_k=1\oplus g_k$.

Since $g_3=2$, $g_4=6$, we have $h_3=1\oplus g_3=3$ and $h_4=1\oplus g_4=7$. Therefore, two consecutive vertices of the inserted path are
$128+h_3=131$, $128+h_4=135$. Hence $\{131,135\}$ is an internal edge of the
inserted path, and its image $e=\{150,182\}$ is an internal edge of $A_8$.

We next determine the residual endpoint edge set. The only degree-one vertices of $A_8$ are the four vertices appearing in $\operatorname{End}(A_8)=\{\{14,38\},\{46,54\}\}$.
Among the six possible pairs of these four vertices, exactly three pairs have Hamming distance one: $\{14,46\}$, $\{38,46\}$ and $\{38,54\}$. None of these three edges belongs to $A_8$. Consequently, $R(A_8)=\{\{14,46\},\{38,46\},\{38,54\}\}$.

The edge $e$ lies on the path with endpoints
$14$ and $38$, and deleting $e$ splits this
path into two paths. Therefore, the resulting endpoint pairs are $\operatorname{End}(A_8\setminus\{e\})=\{\{14,150\},\{38,182\},\{46,54\}\}$. The configuration of $A_8$ is shown in Fig. \ref{fig:A-8W-9}(a), with the relevant vertices labelled.

This proves all claimed properties of $A_8$.
\end{proof}

We use the following two Hamiltonian cycles $C_0$ and $C_1$ of $Q_6$.
\[
\begin{aligned}
C_0=\langle&
14,15,13,12,8,9,11,10,2,3,1,0,4,5,7,6,22,23,21,20,16,17,19,18,26,27,\\
&25,24,28,29,31,30,62,63,61,60,56,57,59,58,50,51,49,48,52,53,55,54,\\
&38,39,37,36,32,33,35,34,42,43,41,40,44,45,47,46\rangle,
\end{aligned}
\]
\[
\begin{aligned}
C_1=\langle&
10,11,9,8,12,13,15,14,6,7,5,4,0,1,3,2,18,19,17,16,20,21,23,22,30,31,\\
&29,28,24,25,27,26,58,59,57,56,60,61,63,62,54,55,53,52,48,49,51,50,\\
&34,35,33,32,36,37,39,38,46,47,45,44,40,41,43,42\rangle.
\end{aligned}
\]

\begin{lemma}\label{lem:A2}
There exists a three-path spanning linear forest
$W_9\subseteq E(Q_9)$ such that
$\operatorname{End}(W_9)=
\{\{14,294\},\{46,54\},\{302,310\}\}$, and
$R(W_9)=\{\{14,46\},\{46,302\},\{54$, $310\},\{294,302\},\{294,310\}\}$. 
\end{lemma}

\begin{proof}
We construct $W_9$ from the two Hamiltonian cycles
$C_0$ and $C_1$ defined above. First, remove the two edges $\{14,46\}$ and $\{54,55\}$ from $C_0$, and denote the resulting spanning linear forest by $F'_0$. Then $\operatorname{End}(F'_0)=\{\{14,55\},\{46,54\}\}$. Similarly, remove the two edges $\{38,46\}$ and $ \{54,55\}$ from $C_1$, and denote the resulting spanning linear forest by $F'_1$. Then $\operatorname{End}(F'_1)=\{\{38,55\},\{46,54\}\}$.

We apply two successive path-forest expansions to
each of $F'_0$ and $F'_1$. The first expansion is performed at the internal edge $\{0,1\}$ in $Q_6$. The edge $\{0,4\}$ belongs to both original forests and remains internal after this expansion. We perform the second expansion at this preserved edge.

We denote the resulting forests in $Q_8$ by $F_0$ and $F_1$, respectively. By Property \ref{prop:expansion}, the endpoint pairs of all paths remain unchanged. Therefore, $\operatorname{End}(F_0)=\{\{14,55\},\{46,54\}\}$, and $\operatorname{End}(F_1)=\{\{38,55\},\{46,54\}\}$.

We embed $F_0$ into the lower half of $Q_9$ and
embed $F_1$ into the upper half of $Q_9$. The
vertices of the upper copy are translated by
$256$, so that $38,46,54,55$ are mapped respectively to $294,302,310,311$.

Now add the edge $\{55,311\}$. This edge joins the path component of $F_0$ ending
at $55$ with the corresponding path component in
the upper copy of $F_1$ beginning at $311$.
Hence these two components merge into a single path whose endpoints are $14$ and $294$. The remaining two path components have endpoint
pairs $\{46,54\}$ and $\{302,310\}$. Consequently,
$\operatorname{End}(W_9)=
\{\{14,294\},\{46,54\},\{302,310\}\}$.

It remains to determine the residual endpoint edge set. The degree-one vertices of $W_9$ are exactly $14,294,46,54,302,310$. The adjacency of two endpoint vertices in $Q_9$ is equivalent to their XOR being a power of two. Checking the $\binom{6}{2}$ endpoint pairs using
the criterion $u\oplus v=2^j$ gives exactly
\[
\{14,46\},\quad
\{46,302\},\quad
\{54,310\},\quad
\{294,302\},\quad
\{294,310\}.
\]
None of these edges belongs to $W_9$. Hence
\[
R(W_9)
=
\{\{14,46\},\{46,302\},
\{54,310\},\{294,302\},\{294,310\}\}.
\]
The three paths of $W_9$ are shown in Fig. \ref{fig:A-8W-9}(b). Its three prescribed paths have endpoint pairs $\{14,294\}$, $\{46,54\}$, and $\{302,310\}$, respectively.
\end{proof}


The following theorem of Dvo\v{r}\'ak provides a useful criterion for extending prescribed edges to Hamiltonian cycles in hypercubes.

\begin{theorem}[\cite{Dvorak2005}]
\label{thm:dvorak}
For $n\ge 2$ and $F\subseteq E(Q_n)$ with
$|F|\le 2n-3$, the set $F$ is contained in
a Hamiltonian cycle of $Q_n$ if and only if
$F$ is a linear forest.
\end{theorem}

The following lemma provides a deterministic construction for our reduction based on Theorem~\ref{thm:dvorak}, with running time polynomial in the number of vertices of the hypercube.

\begin{lemma}
\label{lem:small-matching-extension}
Let $s\ge 3$, and let $M$ be a matching of $Q_s$ with $|M|\leq3$. There is a deterministic algorithm that constructs the edge list of a Hamiltonian cycle of $Q_s$ containing $M$ in $O(2^s s^{O(1)})$ time.
\end{lemma}

\begin{proof}
For $s\le5$, the required cycle exists by Theorem~\ref{thm:dvorak} and can be found by exhaustive search in bounded time. Suppose that $s\ge6$. Extend the set of dimensions of the edges in $M$ to a set $I$ of four dimensions. Delete all edges whose dimensions lie outside $I$. The remaining graph is a disjoint union of $Q_4$-subcubes and contains every edge of $M$. These subcubes are indexed by $Q_{s-4}$.

Fix a dimension $h\in I$ not used by $M$, and choose an edge $f_0=uv$ of dimension $h$ in one subcube. Denote its corresponding copy in each subcube $H$ by $f_0^H=u_Hv_H$. Since $f_0^H\notin M$ and $M\cap E(H)$ is a matching, $(M\cap E(H))\cup\{f_0^H\}$ is a linear forest with at most four edges. By Theorem~\ref{thm:dvorak}, it is contained in a Hamiltonian cycle $C_H$, which can be found by exhaustive search on sixteen vertices.

List the subcubes in cyclic binary reflected Gray order. Delete $f_0^H$ from each $C_H$, obtaining a spanning path with endpoints $u_H,v_H$. Around this cyclic order, join consecutive paths alternately by the corresponding cross edges $u_Hu_{H'}$ and $v_Hv_{H'}$. Since the number of subcubes is even, each path endpoint receives exactly one cross edge, and the paths form a single Hamiltonian cycle containing $M$. Making all choices in a fixed order gives a deterministic construction in $O(2^s s^{O(1)})$ time.
\end{proof}

\section{Uniform addresses and prescribed joins}

\subsection{Address coordinates and logical locations}

The coordinates of the host hypercube $Q_n$ are split into an address block of size $a$ and a local block of size $d$, where $n=a+d$. Accordingly, $Q_n=Q_a\square Q_d$.

The address cube $Q_a$ must accommodate both the positions of all gadgets and the routing paths that connect them. We represent each vertex of $Q_a$ by a full address $(u,z,\lambda)\in V(Q_a)$, consisting of three coordinate blocks: a base-address block $u\in\{0,1\}^B$, a route tag $z\in\{0,1\}^{r_b}$, and a two-bit lane label $\lambda\in\{0,1\}^2$. Thus, the dimension of the address cube is $a=B+r_b+2$. The route tag $z$ distinguishes the $H=3m+t$ logical connections in the construction, comprising $3m$ literal occurrences and $t$ terminal variable connections; hence, $r_b=\lceil\log_2(H+1)\rceil$. When $z=0$, the symbol $0$ denotes the all-zero $r_b$-bit string.

The base address specifies the logical location of a gadget and consists of five consecutive blocks: a clause identifier of length $3h_c$, a variable identifier of length $h_v$, a six-bit marker block, an auxiliary Gray-code block of length $h_\omega$, and a single-bit terminal flag $\tau$. Thus, $B=3h_c+h_v+6+h_\omega+1$, where $h_c=\max\{1,\lceil\log_2 m\rceil\}$, $h_v=\max\{1,\lceil\log_2 t\rceil\}$, and $h_\omega=\lceil\log_2(2t+1)\rceil$. For integers $0\le x<2^h$, let $\operatorname{bin}_h(x)$ denote the standard $h$-bit binary representation of $x$, and let $\rho_h(x)$ denote the concatenation of three identical copies of $\operatorname{bin}_h(x)$, having total length $3h$. For $0\le j\le t$, let $\omega_j$ be the $h_\omega$-bit reflected Gray encoding of $2j$. The words $\omega_0,\ldots,\omega_t$ are distinct and have even Hamming weight. Consequently, $d_H(\omega_j,\omega_{j'})\ge 2$ whenever $j\ne j'$, and $d_H(\omega_{i-1},\omega_i)=2$ for every $1\le i\le t$. 

For each clause $C_q$, define
$c_q=(\rho_{h_c}(q-1),0^{h_v},000000,0^{h_\omega},0)$. We reserve the first three coordinates of the six-bit marker block for the three literal positions. Put $\sigma_1=100000$, $\sigma_2=010000$, and $\sigma_3=001000$, and define the literal attachment addresses by
$\ell_{q,k}=(\rho_{h_c}(q-1),0^{h_v},\sigma_k,0^{h_\omega},0)$ for
$k\in\{1,2,3\}$. The base address $\ell_{q,k}$ identifies the location of the $k$-th literal occurrence of clause $C_q$. Specifically, the
corresponding variable route contains the vertex
$(\ell_{q,k},0,00)$ in $Q_a$.


For the variable identifiers, let $\gamma_i$ be the $h_v$-bit reflected Gray encoding of $i-1$. Thus consecutive variables have Gray identifiers at Hamming distance one. Define
$s_i=(0^{3h_c},\gamma_i,111000,\omega_{i-1},1)$ and $t_i=(0^{3h_c},\gamma_i,111000,\omega_i,0)$. For $1\le i<t$, define the bridge address $b_i=(0^{3h_c},\gamma_{i+1},111000,\omega_i,0)$. Thus $t_i$ and $b_i$ differ only in the variable identifier block, while $b_i$ and $s_{i+1}$ differ only in the terminal flag. These two adjacencies will later be used to attach the bridge between the constructions for variables $x_i$ and $x_{i+1}$.

\subsection{Variable routes in the address cube}

Fix a variable $x_i$, and list all literal occurrences $(q,k)$ satisfying $i(q,k)=i$ in increasing order of their clause indices. If these occurrences are $(q_{i,1},k_{i,1}),\ldots(q_{i,\nu_i},k_{i,\nu_i})$, consider the logical sequence $s_i,\,
\ell_{q_{i,1},k_{i,1}},\ldots,
\ell_{q_{i,\nu_i},k_{i,\nu_i}},\,
t_i$. Each consecutive pair in this sequence determines one logical connection. Since $\sum_i\nu_i=3m$, the total number of such connections is
$\sum_i(\nu_i+1)=3m+t=H$. Assign a distinct nonzero
$r_b$-bit route tag to every logical connection.

Let $u,v$ be any consecutive pair of base addresses along the logical sequence of a variable $x_i$ (that is, $u, v \in \{s_i, \ell_{q_{i,1},k_{i,1}}, \dots, t_i\}$). A logical connection between base addresses $u$ and $v$ with route tag $z$ is represented by a path in the address cube. We call this path a \emph{leg}. It is given by
\begin{equation*}
\begin{aligned}
(u,0,00)&\rightarrow(u,0,10)\rightarrow(u,z,10)
\rightarrow(u,z,11)\\
&\rightarrow(v,z,11)\rightarrow(v,z,01)
\rightarrow(v,0,01)\rightarrow(v,0,00).
\end{aligned}
\end{equation*}
If two consecutive displayed addresses differ in more than one coordinate, the arrow between them denotes the shortest cube path obtained by flipping the differing coordinates one at a time, in increasing order of their indices.

Expanding each multi-coordinate transition of a leg into the specified sequence of single-coordinate flips and concatenating the legs in the order of the logical sequence produces the routed address path $v_{i,0},v_{i,1},\ldots,v_{i,L_i}$ for variable $x_i$, where consecutive vertices are adjacent in $Q_a$. Since each leg from $u$ to $v$ starts at $(u,0,00)$ and ends at $(v,0,00)$, and the logical sequence starts at $s_i$ and ends at $t_i$, we have $v_{i,0}=(s_i,0,00)$, $v_{i,L_i}=(t_i,0,00)$.

The internal junctions between consecutive legs are precisely the logical locations $(\ell_{q,k},0,00)$ corresponding to the literal occurrences of $x_i$. The other vertices strictly inside the expanded legs are auxiliary routing vertices, each having a nonzero lane label. Consequently, the vertices on the route with route tag $0$ and lane label $00$ are precisely its logical locations: the initial address $(s_i,0,00)$, the specified literal-occurrence addresses, and the terminal address $(t_i,0,00)$. By the definitions of $s_i$ and $t_i$, the auxiliary words $\omega_{i-1}$ and $\omega_i$ differ in exactly two coordinates, while their terminal flags 1 and 0 differ in one coordinate. Hence $d_H(s_i,t_i)=3$. By the endpoint identities established above, $d_H(v_{i,0},v_{i,L_i})=3$. Since every edge of the hypercube changes exactly one coordinate, the length of any path between two vertices has the same parity as their Hamming distance. Therefore,
$L_i\equiv d_H(v_{i,0},v_{i,L_i})\equiv d_H(s_i,t_i)\equiv 3\equiv 1\pmod 2$. In particular, $L_i$ is odd. 

Note that each literal occurrence $(q,k)$ of variable $x_i$ is placed at an intermediate vertex $v_{i,p(q,k)} = (\ell_{q,k}, 0, 00)$ for a unique index $p(q,k) \in \{1, \dots, L_i-1\}$. Here, the base address $\ell_{q,k}$ shares the clause identifier $\rho_{h_c}(q-1)$ with the clause base address $c_q$, differing from $c_q$ only in the $k$-th marker coordinate. The vertex $v_{i,p(q,k)}$ is incident with the two consecutive route edges $\{v_{i,p(q,k)-1},v_{i,p(q,k)}\}$ and $\{v_{i,p(q,k)},v_{i,p(q,k)+1}\}$. Since the full addresses $(\ell_{q,k},0,00)$ and $(c_q,0,00)$ are adjacent in $Q_a$, this vertex provides the attachment position for the clause subcube corresponding to $C_q$.

\subsection{Local labels and spanning forests}
To fix the dimension of the local subcube $Q_d$, let $L_{\max} = \max_{1 \le i \le t} L_i$, put $r = \lceil\log_2 L_{\max}\rceil$, and set $d = 8 + r + 3h_v$. For each address $\alpha \in V(Q_a)$, let $Q_d(\alpha)$ denote the corresponding $d$-dimensional subcube of $Q_n$ at address $\alpha$. Whenever $\alpha\beta \in E(Q_a)$, the identity matching between $V(Q_d(\alpha))$ and $V(Q_d(\beta))$ formed by edges $\{(\alpha, y), (\beta, y)\}$ for all $y \in V(Q_d)$ defines the set of \emph{cross-subcube edges} between them.

The $d$ local coordinates consist of a $3h_v$-bit
variable prefix, an $r$-bit route-link code, and an
eight-bit local port label. For $0\le j<L_{\max}$, let $\theta_j$ be the $r$-bit reflected Gray encoding of $j$. For $1\le i\le t$, $0\le j<L_i$, and $\zeta\in\{0,\ldots,255\}$, define
\[
\Lambda(i,j,\zeta)
=
\bigl(
\rho_{h_v}(i-1),
\theta_j,
\operatorname{bin}_8(\zeta)
\bigr)
\in\{0,1\}^d.
\]
The parameter $j$ refers to the route link joining
$v_{i,j}$ and $v_{i,j+1}$, while $\zeta$ is an
eight-bit local port label. In particular, the values $14,38,46,54,150,$ and $182$ used below refer only to the last eight local coordinates.

We now place a spanning linear forest inside every local subcube associated with the logical construction. The linear forests from Lemmas \ref{lem:A1} and \ref{lem:A2} serve as low-dimensional templates. By repeated application of the path-forest expansion operation, each copy of $A_8$ or $W_9$ is enlarged to a spanning linear forest of the required local $d$-dimensional subcube. The expansion preserves the relevant endpoint structure and, for the $A_8$ template, the specified internal edge $\{150,182\}$.

At the initial and terminal route positions
$v_{i,0}$ and $v_{i,L_i}$, place an expanded copy of the two-path forest $A_8$. We refer to these as the \emph{terminal $A_8$-subcubes} of variable $x_i$. At every internal route position $v_{i,p}$ with $0<p<L_i$, place an expanded copy of the three-path forest $W_9$. Since the consecutive Gray words $\theta_{p-1}$ and $\theta_p$ differ in exactly one coordinate, we identify the ninth coordinate of the $W_9$ template with this unique differing coordinate. All other coordinates of the $r$-bit route-position block remain fixed within this copy of $W_9$. After this identification, the side of the $W_9$-subcube facing route link $(i,p-1)$ has Boolean endpoints with local labels $\Lambda(i,p-1,46)$ and $\Lambda(i,p-1,54)$, while the side facing route link $(i,p)$ has Boolean endpoints with local labels $\Lambda(i,p,46)$ and $\Lambda(i,p,54)$.

For each route link $j$, let $\chi_j$ denote its common-channel port label, defined by $\chi_j=14$ when $j$ is even and $\chi_j=38$ when $j$ is odd. Thus, the common-channel port labels alternate between $14$ and $38$ along each variable route. At an internal route position $v_{i,p}$, orient the $W_9$-subcube so that its common-channel endpoint on the side of link $p-1$ has local label $\Lambda(i,p-1,\chi_{p-1})$, while its endpoint on the side of link $p$ has local label $\Lambda(i,p,\chi_p)$. Orient the terminal $A_8$-subcubes consistently with their unique adjacent route links. Consequently, the endpoint of the subcube at $v_{i,j}$ facing $v_{i,j+1}$ and the endpoint of the subcube at $v_{i,j+1}$ facing $v_{i,j}$ both have the local label $\Lambda(i,j,\chi_j)$.
For every route link $j$, prescribe the cross-subcube edge
\[
\bigl\{
(v_{i,j},\Lambda(i,j,\chi_j)),
(v_{i,j+1},\Lambda(i,j,\chi_j))
\bigr\}.
\]
Thus, every route link has exactly one prescribed common-channel edge, whose port label alternates between $14$ and $38$. The two parallel cross-subcube edges with local labels $\Lambda(i,j,46)$ and $\Lambda(i,j,54)$ are not prescribed; they will serve as the two Boolean candidates in a Hamiltonian extension.

We next define the clause constructions. For $b\in\{0,1\}$, put $\zeta_0=54$ and $\zeta_1=46$. For a literal occurrence $(q,k)$, let $i=i(q,k)$ and $p=p(q,k)$. In the local subcube $Q_d((c_q,0,00))$, select the edge
\[
e_{q,k}=\left\{\Lambda(i,p-1,\zeta_{b(q,k)}),\Lambda(i,p,\zeta_{b(q,k)})\right\}.
\]
The two endpoints of $e_{q,k}$ have the same variable prefix and the same local port label, and differ only in the consecutive route-position codes $\theta_{p-1}$ and $\theta_p$. Since these two Gray words differ in exactly one coordinate, $e_{q,k}$ is an edge of the local $Q_d$-subcube. The edge $e_{q,k}$ is chosen as the cut edge associated with the literal occurrence $(q,k)$: after it is removed from the clause Hamiltonian cycle, its two endpoints become degree-one vertices corresponding to the two sides of the $W_9$-subcube placed at the literal route position $v_{i,p}$. These two exposed clause endpoints may then be joined, through the corresponding cross-subcube edges, to the two matching endpoints of that $W_9$-subcube. The cross-subcube edges are left unprescribed and form the optional external connection associated with $(q,k)$. 

The three cut edges $e_{q,1},e_{q,2},e_{q,3}$ form an induced matching by Lemma \ref{lem:address_geometry}(4). Apply Lemma \ref{lem:small-matching-extension} in the local subcube at $(c_q,0,00)$ to construct a Hamiltonian cycle $C_q^{\mathrm{loc}}$ containing all three cut edges. Define the clause forest by $F_q=C_q^{\mathrm{loc}}-\{e_{q,1},e_{q,2},e_{q,3}\}$. Then $F_q$ is a spanning linear forest consisting of three paths, with six degree-one vertices given by the endpoints of the three deleted edges. We call $F_q$ the three-cut clause forest associated with $C_q$. The resulting local structure is illustrated in Fig. \ref{fig:A-8W-9}(c), where the three deleted edges are indicated by orange dashed lines. Note that $e_{q,1},e_{q,2},e_{q,3}$ are not included in the prescribed forest. In fact, they remain available for use in a Hamiltonian extension. For any selected index $k\in\{1,2,3\}$, restoring the other two cut edges gives $F_q+\{e_{q,k'}:k'\ne k\}=C_q^{\mathrm{loc}}-\{e_{q,k}\}$, which is a  Hamiltonian path whose endpoints are the two endpoints of $e_{q,k}$. 

We now specify the external connections associated with $e_{q,1},e_{q,2}$ and $e_{q,3}$. The endpoints of $e_{q,k}$ are degree-one vertices of the clause forest $F_q$. Since the outer addresses $(c_q,0,00)$ and $v_{i,p}=(\ell_{q,k},0,00)$ are adjacent in $Q_a$, these two endpoints can be connected to the corresponding vertices of the $W_9$-subcube in $Q_d(v_{i,p})$ by the cross-subcube edges
\[
\begin{aligned}
g_{q,k}^{-}
&=\left\{\bigl((c_q,0,00),\Lambda(i,p-1,\zeta_{b(q,k)})\bigr),\right.
\left.\bigl(v_{i,p},\Lambda(i,p-1,\zeta_{b(q,k)})\bigr)\right\},\\
g_{q,k}^{+}
&=\left\{\bigl((c_q,0,00),\Lambda(i,p,\zeta_{b(q,k)})\bigr),\right.
\left.\bigl(v_{i,p},\Lambda(i,p,\zeta_{b(q,k)})\bigr)\right\}.
\end{aligned}
\]
Neither $g_{q,k}^{-}$ nor $g_{q,k}^{+}$ belongs to the prescribed forest. Together, these two edges provide the optional external connection between the clause subcube and the variable subcube at the literal position $v_{i,p}$.

Finally, consider the bridge between variables $x_i$ and $x_{i+1}$, where $1\le i<t$. In the terminal $A_8$-subcube of $x_i$, delete the internal edge corresponding to the template edge $\{150,182\}$, whose endpoints have local labels $\Lambda(i,L_i-1,150)$ and $\Lambda(i,L_i-1,182)$. In the initial $A_8$-subcube of $x_{i+1}$, delete the corresponding internal edge with endpoints $\Lambda(i+1,0,150)$ and $\Lambda(i+1,0,182)$.

Inside the bridge subcube $Q_d((b_i,0,00))$,
define the two edges
\[
\begin{aligned}
e_i^-&=\{
\Lambda(i,L_i-1,150),
\Lambda(i,L_i-1,182)
\},\\
e_i^+&=\{
\Lambda(i+1,0,150),
\Lambda(i+1,0,182)
\}.
\end{aligned}
\]
Hence, $e_i^-$ and $e_i^+$ form an induced matching in the bridge subcube by Lemma \ref{lem:address_geometry}(4). By Lemma \ref{lem:small-matching-extension}, there exists a Hamiltonian cycle $C_i^{\mathrm{br}}$ in this subcube containing both edges. Deleting these two edges from $C_i^{\mathrm{br}}$ yields a spanning linear forest consisting of exactly two vertex-disjoint paths, with the four endpoints given by the endpoints of $e_i^-$ and $e_i^+$. We call this forest the \emph{two-cut bridge forest} (see Fig. \ref{fig:A-8W-9}(d)). We then prescribe the four cross-subcube edges joining these four degree-one vertices to the four newly exposed degree-one vertices of the adjacent terminal $A_8$-subcubes, along the outer cube edges $\{(t_i,0,00), (b_i,0,00)\}$ and $\{(b_i,0,00), (s_{i+1},0,00)\}$.

The outer addresses assigned to the variable-route
positions, clause gadgets, and bridge gadgets are
called \emph{occupied addresses}, and all remaining
outer addresses are called \emph{unoccupied addresses}. Accordingly, the corresponding local subcubes are called \emph{occupied subcubes} and \emph{unoccupied subcubes}, respectively.

Let $G_{\mathrm{occ}}(\Phi)$ denote the graph whose vertices are all vertices in the occupied local subcubes and whose edges are precisely the prescribed edges specified within and between these subcubes. The Boolean candidate edges and the optional clause-variable connection edges are not included in $G_{\mathrm{occ}}(\Phi)$. Lemma~\ref{lem:complete} shows how a satisfying assignment determines additional edges that complete $G_{\mathrm{occ}}(\Phi)$ to a Hamiltonian cycle on the occupied vertices.

\begin{figure}[t]
    \centering
    \includegraphics[
        width=0.99\textwidth,
        height=0.88\textheight,
        keepaspectratio
    ]{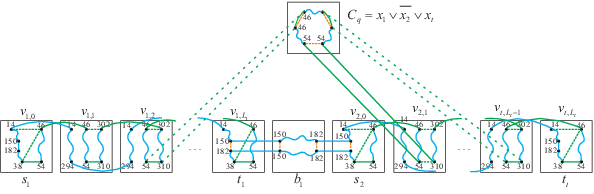}
    \caption{A schematic Hamiltonian-cycle construction for the clause $C_q=(x_1\vee\neg x_2\vee x_t)$ under the assignment $(x_1,x_2,x_t)=(0,0,0)$. Blue solid edges denote prescribed edges, whereas green solid edges denote the optional edges selected under the given assignment. The Hamiltonian cycle is the union of all blue and green solid edges shown in the figure.}
    \label{fig:combined-clause-variable-bridge}
\end{figure}

Fig. \ref{fig:combined-clause-variable-bridge}
illustrates this construction for the representative
clause $C_q=(x_1\vee \neg x_2\vee x_t)$ and the
assignment $(x_1,x_2,x_t)=(0,0,0)$, showing how the
prescribed edges and the selected optional edges
combine to form the Hamiltonian cycle.

\subsection{Address separation and validity of the prescribed connections}

\begin{lemma}\label{lem:address_geometry}
The address assignments defined above have the following properties.
\begin{enumerate}
    \item For each variable $x_i$, the route $v_{i,0},\ldots,v_{i,L_i}$ is a simple path in $Q_a$. Two distinct legs are vertex-disjoint unless they are consecutive legs of the same variable route; in that case, they share exactly one vertex, namely the literal address $(\ell_{q,k},0,00)$ at which they are concatenated.

    \item Among the occupied addresses of $Q_a$, each clause address $(c_q,0,00)$ is adjacent exactly to its three literal addresses $(\ell_{q,k},0,00)$, $k\in\{1,2,3\}$. Each bridge base address $(b_i,0,00)$ is adjacent precisely to $(t_i,0,00)$ and $(s_{i+1},0,00)$.

    \item All prescribed common-channel edges, all prescribed bridge-attachment edges, and all optional clause-variable connection edges belong to $E(Q_n)$.

    \item The three cut edges in each clause subcube form an induced matching of size three, and the two cut edges in each bridge subcube form an induced matching of size two.
\end{enumerate}
\end{lemma}

\begin{proof}
(1) Each leg is a simple path, since its successive coordinate paths flip distinct coordinates without repetition and occupy different lanes. For distinct legs, their segments in lane $10$ are disjoint because each such segment has a fixed initial base address, and distinct legs have distinct initial base addresses. Similarly, their segments in lane $01$ are disjoint because each such segment has a fixed terminal base address, and distinct legs have distinct terminal base addresses. Their segments in lane $11$ are disjoint because each leg is assigned a distinct nonzero route tag.

(2) We verify the asserted adjacencies among all occupied addresses of $Q_a$.
First consider the logical addresses with route tag $0$ and lane label $00$.
For each clause $C_q$, the three literal addresses
$\ell_{q,1},\ell_{q,2},\ell_{q,3}$ differ from $c_q$ in exactly one marker
coordinate, so they are adjacent to $c_q$. If $q'\ne q$, then
\[
d_H\bigl(\rho_{h_c}(q-1),\rho_{h_c}(q'-1)\bigr)
=3d_H\bigl(\operatorname{bin}_{h_c}(q-1),
\operatorname{bin}_{h_c}(q'-1)\bigr)\ge 3.
\]
Thus no other clause address or literal address is adjacent to $c_q$. Moreover, the marker block of every variable address $s_i,t_i$ and every bridge address $b_i$ is $111000$, which differs from the marker block $000000$ of $c_q$ in three coordinates. Hence the only occupied logical addresses adjacent to $c_q$ are its three literal addresses.

Now fix a bridge address $b_i$. Its marker block $111000$ differs from the marker block of each clause address in three coordinates and from
the marker block $\sigma_k$ of each literal address in two coordinates. For any terminal variable address $t_j$, the definitions give $d_H(b_i,t_j)=d_H(\gamma_{i+1},\gamma_j)+d_H(\omega_i,\omega_j)$, which is one when $j=i$, and is at least two when $j\ne i$. Similarly, $d_H(b_i,s_j)
=d_H(\gamma_{i+1},\gamma_j)
+d_H(\omega_i,\omega_{j-1})+1$, which is one when $j=i+1$, and is at least three otherwise.
Finally, if $j\ne i$, then $d_H(b_i,b_j)\ge d_H(\omega_i,\omega_j)\ge 2$. Consequently, among the logical addresses, $b_i$ is adjacent
exactly to $t_i$ and $s_{i+1}$.

It remains to consider the auxiliary vertices introduced by the expanded legs. Every such vertex has a nonzero lane label. If it is adjacent to a vertex of the form $(a,0,00)$, then its route tag must be zero and its base address must equal $a$. By the construction of a leg from $u$ to $v$, the only auxiliary vertices with zero route tag are $(u,0,10)$ and $(v,0,01)$. Consequently, an auxiliary vertex can be adjacent to a location of the form $(a,0,00)$ only when that location is an endpoint of its own leg. 

(3) Since $Q_n=Q_a\square Q_d$, two vertices with adjacent outer addresses and identical local coordinates are adjacent in $Q_n$. The common-channel edges use consecutive vertices of a variable address path. The bridge-attachment edges use the outer-address adjacencies established in (2), and the optional clause-variable edges use the adjacency between $c_q$ and $\ell_{q,k}$. Thus all the stated edges belong to $Q_n$.

(4) Each clause cut edge $e_{q,k}$ is an edge of its local $Q_d$ because the consecutive Gray codes $\theta_{p-1}$ and $\theta_p$ differ in exactly one coordinate. The three literals of a clause involve distinct variables, whose triple-repeated variable prefixes have pairwise Hamming distance at least three. Therefore, endpoints belonging to distinct clause cut edges are nonadjacent, and the three edges form an induced matching.

For a bridge cut edge, the local labels $150$ and $182$ differ in exactly one coordinate, since $150\oplus182=32$. The two bridge cut edges use distinct variable prefixes at Hamming distance at least three, so their endpoints are pairwise nonadjacent across the two edges. Hence they form an induced matching of size two.
\end{proof}

\subsection{Degree-feasible cross-subcube edges}

We now determine which non-prescribed cross-subcube edges can be used in a Hamiltonian extension without violating the degree-two condition. Let $P$ consist of all prescribed edges, including the edges of the local spanning forests and the prescribed cross-subcube edges. Since every vertex of a Hamiltonian cycle has degree two, a non-prescribed edge joining two vertices of prescribed degree one is called {\em degree-feasible}. For each occupied outer address $\alpha$, define
\[
T_\alpha=\{y\in\{0,1\}^d:
\deg_P((\alpha,y))=1\}.
\]
Thus, $T_\alpha$ records the local coordinates of vertices that can be incident with one additional edge in Hamiltonian completion. If $\alpha\beta\in E(Q_a)$, the degree-feasible cross-subcube edges between $Q_d(\alpha)$ and $Q_d(\beta)$ are precisely
\[
\bigl\{\{(\alpha,y),(\beta,y)\}:y\in T_\alpha\cap T_\beta\bigr\}.
\]
Since the endpoints of every prescribed cross-subcube edge already have prescribed degree two, no such edge belongs to this set.

\begin{lemma}\label{lem:cross_subcube_edges}
For adjacent occupied addresses in $Q_a$, the prescribed and degree-feasible cross-subcube edges are as follows.
\begin{enumerate}
\item Between $v_{i,j}$ and $v_{i,j+1}$, there is
one prescribed common-channel edge corresponding
to the port $\chi_j$ (that is, $14$ if $j$ is even
and $38$ if $j$ is odd), together with two
degree-feasible edges corresponding to the ports
$46$ and $54$.
\item Between $(c_q,0,00)$ and $v_{i,p(q,k)}$, where $i=i(q,k)$, the only degree-feasible edges are $g_{q,k}^{-}$ and $g_{q,k}^{+}$; neither is prescribed.
\item Between $(b_i,0,00)$ and either $(t_i,0,00)$ or $(s_{i+1},0,00)$, there are exactly two prescribed edges and no degree-feasible edges.
\item All other adjacent occupied addresses have no prescribed or degree-feasible cross-subcube edges between them.
\end{enumerate}
\end{lemma}

\begin{proof} (1)--(3) follow directly from the definitions of the local forests and their prescribed connections. For (4), it suffices to consider pairs of route positions $v_{i,p}$ and $v_{i',q}$ that are not consecutive on the same variable route. If $i\ne i'$, $\rho_{h_v}(i-1)$ and $\rho_{h_v}(i'-1)$ differ in at least three coordinates. Hence there is no local coordinate $y$ for which both $(v_{i,p},y)$ and $(v_{i',q},y)$ have degree one in $P$. Consequently, no degree-feasible cross-subcube edge can join the two corresponding local subcubes.

Suppose now that $i=i'$ and $p<q$. Since the two positions are not consecutive, $q-p\ge 2$. Thus, the local labels of the degree-one vertices in
$Q_d(v_{i,p})$ are of the form $\Lambda(i,0,\zeta)$
if $p=0$, of the forms $\Lambda(i,p-1,\zeta)$ and
$\Lambda(i,p,\zeta)$ if $0<p<L_i$, and of the form
$\Lambda(i,L_i-1,\zeta)$ if $p=L_i$, for the
corresponding port labels specified above. Since $q-p\ge 2$, the two positions have no common
route link. 

Therefore, in either case, no degree-feasible non-prescribed cross-subcube edge can join the two corresponding local subcubes. By construction, no prescribed cross-subcube edge is assigned between these pairs either.
\end{proof}

\subsection{Acyclicity of the prescribed edges on occupied subcubes}

\begin{lemma}\label{lem:occupied_forest}
The graph $G_{\mathrm{occ}}(\Phi)$ is a spanning linear forest with exactly
\[
c_0=2\sum_{i=1}^{t}L_i+t+3m
\]
components.
\end{lemma}

\begin{proof}
Recall that $L_i$ is the length of the expanded route $v_{i,0},v_{i,1},\ldots,v_{i,L_i}$ of variable $x_i$. After deleting the edge corresponding to $\{150,182\}$ from each $A_8$-subcube incident with a bridge subcube, the forest has
\[
\begin{aligned}
N_{\mathrm{loc}}
&=4+6(t-1)+3(\sum_{i=1}^t L_i-t)+2(t-1)+3m\\
&=3\sum_{i=1}^t L_i+5t-4+3m
\end{aligned}
\]
path components. Here, the two end $A_8$-subcubes contribute four paths, the $2(t-1)$ bridge-incident $A_8$-subcubes contribute three paths each, each $W_9$-subcube contributes three paths, each bridge subcube contributes two paths, and each clause subcube contributes three paths.

Add the prescribed common-channel edges in route order. Each edge joins distinct components, extending the common-channel paths along the variable routes without creating a cycle. Since there are $\sum_{i=1}^t L_i$ such edges, the component count decreases by $\sum_{i=1}^t L_i$.

Each bridge subcube consists of two disjoint paths. The four prescribed attachment edges join their endpoints to the two distinct exposed components on each side, extending the paths without creating a cycle. Thus each bridge reduces the number of path components by four.

Consequently, $G_{\mathrm{occ}}(\Phi)$ is acyclic and has
\[
\begin{aligned}
c_0
&=N_{\mathrm{loc}}-\sum_{i=1}^t L_i-4(t-1)\\
&=2\sum_{i=1}^t L_i+t+3m
\end{aligned}
\]
components.

Every vertex has prescribed degree one or two. Hence every component is a path, and $G_{\mathrm{occ}}(\Phi)$ is a spanning linear forest with exactly $c_0$ path components.
\end{proof}

\subsection{Extending the prescribed forest to the unoccupied local subcubes}

The graph $G_{\mathrm{occ}}(\Phi)$ constructed so far spans all vertices in the occupied local subcubes, but it does not yet include the vertices in the unoccupied local subcubes. We now incorporate those remaining local subcubes by path substitutions. 

Whenever an already incorporated subcube $Q_d(\alpha)$ is adjacent to a subcube $Q_d(\beta)$ that has not yet been incorporated, we choose an edge $xy$ of the prescribed forest lying entirely within $Q_d(\alpha)$, and let $x',y'$ be its corresponding vertices in $Q_d(\beta)$. We delete $xy$ and replace it with the two cross-subcube edges $xx'$ and $yy'$, together with a Hamiltonian path of $Q_d(\beta)$ from $x'$ to $y'$. The chosen edge $xy$ must be an internal edge of the prescribed local forest in $Q_d(\alpha)$, so that neither $x$ nor $y$ is an endpoint of a path. In particular, $xy$ is not adjacent to any prescribed common-channel edge or prescribed bridge-attachment edge. The newly incorporated subcube can then be used in further extensions.

To specify the order of these extensions, choose a multi-source breadth-first forest $\mathcal{T}$ of $Q_a$, rooted at all addresses corresponding to the initially occupied subcubes. For each
address $\beta$ corresponding to an unoccupied subcube, define its \emph{parent} to be the neighbor of $\beta$ immediately preceding it on the unique path in $\mathcal{T}$ from the root of its tree to $\beta$. Conversely, the \emph{children} of an address $\alpha$ are all addresses $\beta$ whose parent is $\alpha$. Process the addresses in order of increasing distance from the root set, so that each parent subcube is incorporated before its children. Since $Q_a$ has maximum degree $a$, each address has at most $a$ children.

\begin{lemma}\label{lem:unused_subcubes}
Starting from $G_{\mathrm{occ}}(\Phi)$, the
path-substitution process produces the graph
$G_\Phi=(V(Q_{a+d}),P(\Phi))$, which is a spanning
linear forest of $Q_{a+d}$. Moreover, the pairing
of the degree-one endpoints of $G_{\mathrm{occ}}(\Phi)$ is preserved, and every vertex in an unoccupied subcube has degree two.
\end{lemma}

\begin{proof}
Let $N=2^d$. Each initially occupied subcube contains a spanning linear forest with at most three components, and hence at least $N-3$ prescribed edges.

When choosing an edge for a path substitution, we avoid all endpoints of the local paths before the prescribed cross-subcube edges are added. These include every degree-one vertex of $G_{\mathrm{occ}}(\Phi)$ in the subcube and every vertex incident with a prescribed cross-subcube edge. Since the local forest has at most three components, there are at most six such vertices, each incident with exactly one local prescribed edge. Excluding these edges leaves at least $N-9$ eligible edges. Since $\mathcal{T}$ is a rooted spanning forest of $Q_a$, each address has at most $a$ children. Thus, each local subcube needs at most $a$ pairwise vertex-disjoint eligible edges, one for each child. Since the local forest has maximum degree two, each greedy selection eliminates at most three eligible edges, including the selected edge itself. Consequently, we can select the required $a$ pairwise vertex-disjoint eligible edges whenever $N-9>3(a-1)$.

Since each of the $H$ logical legs contains at least seven edges of $Q_a$, we have
\[
\sum_{i=1}^t L_i\ge 7H,
\qquad
L_{\max}\ge \frac{7H}{t}.
\]
Moreover, $d=3h_v+r+8$ and $2^{h_v}\ge t$. Hence
\[
\begin{aligned}
N
&=256\cdot 2^r\cdot 2^{3h_v}\\
&\ge 256\cdot \frac{7H}{t}\cdot t^3\\
&=1792Ht^2.
\end{aligned}
\]

Since $h_c \le m$, $h_v \le t-1$, $h_\omega \le t$, and $r_b \le H$, the definitions of $a$ and $B$ give $a \le 2H + t + 8 \le 3H + 8$. Combining this bound with $N \ge 1792Ht^2$ and the assumptions $m \ge 1$ and $t \ge 3$, we obtain $N > 3a + 6$, and consequently $N \ge 2a + 4$. Therefore, every initially occupied local subcube has enough eligible edges for all its children in $\mathcal{T}$.

For each newly incorporated subcube, write its
Hamiltonian path as
\[
\langle w_0,w_1,\ldots,w_{N-1}\rangle.
\]
Since its address has at most $a$ children in
$\mathcal{T}$, reserve the following matching of size $a$ for their subsequent incorporations: $w_{2k}w_{2k+1}$, $k=1,\ldots,a$. These edges avoid both endpoints of the path because $N\ge 2a+4$.

Now process $\mathcal{T}$ in the specified order. For each address $\beta$ outside the root set, use the edge reserved for $\beta$ in its parent subcube and perform the path substitution described above. If the selected edge in the parent subcube is $xy$, the edge $x'y'$ in the new subcube corresponds to a Hamiltonian path from $x'$ to $y'$ of $Q_d$, since $d\ge2$.

It therefore preserves acyclicity and the original endpoint pairing. The reserved edges within each parent subcube form a matching, so all substitutions can be carried out in the specified order. Eventually every local subcube is incorporated, and every vertex in a newly incorporated subcube has prescribed degree two. Thus, $G_\Phi$ is a spanning linear forest of $Q_n$ with the required properties.
\end{proof}

Thus, the construction transforms each 3-SAT instance $\Phi$ into a prescribed-edge graph
$G_\Phi=(V(Q_{a+d}),P(\Phi))$ on the hypercube
$Q_{a+d}$, yielding the instance $(a+d,P(\Phi))$ of the Hamiltonian-cycle extension problem. The Boolean candidate edges and the optional clause-variable connection edges are not
included in $P(\Phi)$.

\section{Boolean Consistency and Soundness}


We first show that any Hamiltonian cycle extending $P(\Phi)$ in $Q_{a+d}$ can be reduced to a cycle on the occupied local subcubes by reversing the path substitutions of Section~3.7.

\begin{lemma}\label{lem:contract}
Let $C$ be a Hamiltonian cycle of $Q_{a+d}$ containing $P(\Phi)$. Reversing the path substitutions in decreasing distance from the roots of $\mathcal{T}$ produces a cycle $C_{\mathrm{occ}}$ spanning the occupied vertices and containing every edge of $G_{\mathrm{occ}}(\Phi)$. Moreover,
\[
E(C)\setminus P(\Phi)=E(C_{\mathrm{occ}})\setminus E(G_{\mathrm{occ}}(\Phi)).
\]
\end{lemma}
\begin{proof}
By Lemma \ref{lem:unused_subcubes}, every vertex in an unoccupied local subcube has prescribed degree two in $P(\Phi)$, and hence cannot be incident with any non-prescribed edge of $C$. Consider a leaf $\beta$ of the breadth-first forest $\mathcal{T}$ outside the root set, and let $\alpha$ be its parent. In $Q_d(\beta)$, the cycle $C$ must traverse the entire prescribed Hamiltonian path from $x'$ to $y'$ together with the two prescribed cross-subcube edges $xx'$ and $yy'$. Replacing the subpath $\langle x, x', \dots, y', y \rangle$ of $C$ by the original internal edge $xy$ of $Q_d(\alpha)$ yields a single cycle covering all remaining subcubes and containing the pre-substitution prescribed forest. Repeating this contraction in reverse breadth-first order eliminates all unoccupied subcubes without altering any edges incident to degree-one vertices of $G_{\mathrm{occ}}(\Phi)$, yielding the desired cycle $C_{\mathrm{occ}}$.

At every reversal, the removed path belongs to the current prescribed forest, and the restored edge belongs to the preceding prescribed forest; all completion edges remain unchanged.
\end{proof}

By Lemma \ref{lem:contract}, to analyse a Hamiltonian cycle $C$ containing $P(\Phi)$, it suffices to work with its contracted cycle $C_{\mathrm{occ}}$ on the occupied subcubes. For a variable route $x_i$ and a route link $0 \le j < L_i$, let $\epsilon_{i,j,b}(C) \in \{0,1\}$ be the indicator function denoting whether $C$ (and hence $C_{\mathrm{occ}}$) uses the Boolean cross-subcube edge with local port label $\Lambda(i,j,\zeta_b)$ on that link, where $b \in \{0,1\}$ corresponds to $\zeta_0 = 54$ and $\zeta_1 = 46$. Define $\alpha_{q,k}(C) = 1$ if both $g_{q,k}^{-}$ and $g_{q,k}^{+}$ belong to $E(C)$, and $\alpha_{q,k}(C) = 0$ if neither belongs to $E(C)$.

\begin{lemma}\label{lem:atomic}
Let $C$ be a Hamiltonian cycle of $Q_{a+d}$ containing every edge of $P(\Phi)$. Then, for each literal occurrence $(q,k)$, the two optional cross-subcube edges $g_{q,k}^{-}$ and $g_{q,k}^{+}$ either both belong to $C$, or neither does.
\end{lemma}
\begin{proof}
In the prescribed three-cut clause forest $F_q$, the edge $e_{q,k} = uv$ is removed, leaving $u$ and $v$ with degree one in $G_{\mathrm{occ}}(\Phi)$. In the contracted cycle $C_{\mathrm{occ}}$, every vertex has degree two, so $C_{\mathrm{occ}}$ must supply exactly one additional incident edge for both $u$ and $v$. By Lemmas \ref{lem:address_geometry}(4) and \ref{lem:cross_subcube_edges}, the only degree-feasible edges incident to $u$ and $v$ are the internal edge $e_{q,k}$ and the external cross-subcube edges $g_{q,k}^{-}$ and $g_{q,k}^{+}$. If $C$ uses $g_{q,k}^{-}$ at $u$, then $e_{q,k}$ cannot be used, forcing $C$ to use $g_{q,k}^{+}$ at $v$. Conversely, if neither external edge is used, $C$ must restore $e_{q,k}$.
\end{proof}

For each $i \in \{1,\dots,t\}$, set
\[
\rho_i = \begin{cases} 0, & i=1, \\ 2, & 2 \le i \le t. \end{cases}
\]
For each variable $x_i$ and each $0 \le j < L_i$, define $S_{i,j} = \bigcup_{r=0}^{j} V(Q_d(v_{i,r}))$.

\begin{lemma}\label{lem:L6}
Let $C$ be a Hamiltonian cycle containing every edge of $P(\Phi)$, and let $C_{\mathrm{occ}}$ be its contracted cycle from Lemma~\ref{lem:contract}. Then the number of edges of $C_{\mathrm{occ}}$ having exactly one endpoint in $S_{i,j}$ is
\begin{equation}\label{eq:4.1}
1 + \epsilon_{i,j,0}(C) + \epsilon_{i,j,1}(C) + \rho_i + 2 \sum_{(q,k):\, i(q,k)=i,\, p(q,k) \le j} \alpha_{q,k}(C).
\end{equation}
Consequently, exactly one of the two Boolean candidates is used on each route link.
\end{lemma}
\begin{proof}
Fix a variable $x_i$ and an index $0 \le j < L_i$. We count the edges of $C_{\mathrm{occ}}$ with exactly one endpoint in $S_{i,j}$. Since $C_{\mathrm{occ}}$ is supported entirely on the occupied local subcubes and contains $G_{\mathrm{occ}}(\Phi)$, no edges incident with unoccupied subcubes are present in $C_{\mathrm{occ}}$, and every non-prescribed edge of $C_{\mathrm{occ}}$ is identical to that of $C$.

Among the prescribed common-channel edges along the route of $x_i$, only the edge on link $j$, joining $Q_d(v_{i,j})$ and $Q_d(v_{i,j+1})$, has exactly one endpoint in $S_{i,j}$. It therefore contributes $1$. The two other possible edges on this link are the Boolean candidates corresponding to the ports $46$ and $54$. Since $C$ and $C_{\mathrm{occ}}$ share the same non-prescribed edges, their contributions are $\epsilon_{i,j,0}(C)$ and $\epsilon_{i,j,1}(C)$, respectively.

If $i > 1$, the two bridge-attachment edges at the initial subcube of $x_i$ have exactly two endpoints in total in $S_{i,j}$, one from each edge. Thus, they contribute $\rho_i = 2$. For $i = 1$, there is no such bridge, and $\rho_1 = 0$.

Finally, consider a literal occurrence $(q,k)$ of $x_i$ at route position $p(q,k) \le j$. Its two optional clause-variable connection edges join a subcube in $S_{i,j}$ to the corresponding clause subcube outside $S_{i,j}$. By the definition of $\alpha_{q,k}(C)$, these two edges contribute $2\alpha_{q,k}(C)$. Occurrences with $p(q,k) > j$ contribute no edges to this count.

Consequently, adding these contributions gives precisely the expression in~\eqref{eq:4.1}. Since $C_{\mathrm{occ}}$ is a cycle, the number of its edges with exactly one endpoint in $S_{i,j}$ is even. Hence the number counted in~\eqref{eq:4.1} is even, which yields
\[
\epsilon_{i,j,0}(C) + \epsilon_{i,j,1}(C) \equiv 1 \pmod 2.
\]
Since both indicators belong to $\{0,1\}$, their sum is exactly $1$. Thus, $C$ uses exactly one of the two Boolean candidates on each route link.
\end{proof}

\begin{lemma}\label{lem:4.4}
Let $C$ be a Hamiltonian cycle containing every edge of $P(\Phi)$. For each internal route position $v_{i,p}$, the Boolean candidate edges selected by $C$ on its two incident route links have the same local port label in $\{46,54\}$. Moreover, if $v_{i,p}$ is the literal socket for $(q,k)$ and $\alpha_{q,k}(C)=1$, then this common label is $\zeta_{1-b(q,k)}$, while the two clause-variable connection edges use $\zeta_{b(q,k)}$.
\end{lemma}
\begin{proof}
In the $W_9$-subcube at $v_{i,p}$, the prescribed forest consists of the common-channel path and two Boolean paths whose endpoint pairs are $\{\Lambda(i,p-1,46), \Lambda(i,p-1,54)\}$ on the side of link $p-1$ and $\{\Lambda(i,p,46), \Lambda(i,p,54)\}$ on the side of link $p$ (corresponding to $\{46,54\}$ and $\{302,310\}$ in Lemma \ref{lem:A2}). Since the two endpoints of the common-channel path already have degree two in $G_{\mathrm{occ}}(\Phi)$, they cannot be incident with any additional edge in $C_{\mathrm{occ}}$.

By Lemma \ref{lem:L6}, $C_{\mathrm{occ}}$ uses exactly one Boolean candidate edge on link $p-1$, say at port $\zeta \in \{46,54\}$, which saturates $\Lambda(i,p-1,\zeta)$ to degree two. Let $\zeta'$ be the complementary port in $\{46,54\}$. Since the Boolean candidate edge at $\Lambda(i,p-1,\zeta')$ on link $p-1$ is not used, the degree-one vertex $\Lambda(i,p-1,\zeta')$ must receive its second incident edge in $C_{\mathrm{occ}}$ either from the internal residual edge set $R(W_9)$ or from an external clause connection. By Lemma \ref{lem:A2} and Lemma \ref{lem:cross_subcube_edges}, the only such edges incident to $\Lambda(i,p-1,\zeta')$ are the internal residual edge $\{\Lambda(i,p-1,\zeta'), \Lambda(i,p,\zeta')\}$ (corresponding to $\{46,302\}$ or $\{54,310\}$) and, when $v_{i,p}$ is the literal socket for $(q,k)$ with $\zeta' = \zeta_{b(q,k)}$, the external edge $g_{q,k}^{-}$.

If $C_{\mathrm{occ}}$ uses the internal residual edge $\{\Lambda(i,p-1,\zeta'), \Lambda(i,p,\zeta')\}$, then the endpoint $\Lambda(i,p,\zeta')$ on the side of link $p$ also reaches degree two. If instead $C_{\mathrm{occ}}$ uses $g_{q,k}^{-}$, then Lemma \ref{lem:atomic} forces $g_{q,k}^{+} \in E(C_{\mathrm{occ}})$, which is incident to $\Lambda(i,p,\zeta')$ and likewise saturates $\Lambda(i,p,\zeta')$ to degree two. In either case, $\Lambda(i,p,\zeta')$ cannot be incident with the Boolean candidate edge on link $p$. Since Lemma \ref{lem:L6} requires $C_{\mathrm{occ}}$ to use a Boolean candidate edge on link $p$, it must use the one incident to the remaining available endpoint $\Lambda(i,p,\zeta)$. Hence links $p-1$ and $p$ select the same port label $\zeta$. Moreover, when $\alpha_{q,k}(C)=1$, both $g_{q,k}^{-}$ and $g_{q,k}^{+}$ belong to $C_{\mathrm{occ}}$ and saturate the endpoints labeled $\zeta_{b(q,k)}$, forcing $\zeta' = \zeta_{b(q,k)}$ and $\zeta = \zeta_{1-b(q,k)}$.
\end{proof}

\begin{lemma}\label{lem:soundness}
Every Hamiltonian extension of $P(\Phi)$ induces a satisfying assignment of $\Phi$.
\end{lemma}

\begin{proof}
Let $C$ be a Hamiltonian cycle containing every edge of the final prescribed set $P(\Phi)$, and let $C_{\mathrm{occ}}$ be its contracted cycle from Lemma \ref{lem:contract}.

By Lemma \ref{lem:L6}, exactly one of the two Boolean candidate edges is used on each route link $0 \le j < L_i$. Moreover, Lemma \ref{lem:4.4} forces the Boolean candidate edges on the two route links incident with each $W_9$-subcube to have the same local port label. Consequently, all route links of a given variable $x_i$ use the same Boolean port. Define the assignment by setting $x_i=0$ if this port is $46$, and $x_i=1$ if it is $54$.

We next show that every activated literal is true under this assignment. Suppose that the occurrence $(q,k)$ of $x_i$ is activated, so that $\alpha_{q,k}(C)=1$. The two corresponding clause-variable connection edges use the port $\zeta_{b(q,k)}$ at the two sides of the $W_9$-subcube. These edges saturate their incident vertices, so the Boolean channel must use the other port, namely $\zeta_{1-b(q,k)}$. Since $\zeta_0=54$ and $\zeta_1=46$, this means that the variable route uses port $46$ when $b(q,k)=0$ and port $54$ when $b(q,k)=1$. Thus $x_i=b(q,k)$, which is precisely the value that makes the literal occurrence $(q,k)$ true.

It remains to show that every clause has an activated occurrence. Fix a clause $C_q$ and suppose that none of its occurrences is activated. The contracted cycle $C_{\mathrm{occ}}$ contains the original clause forest $F_q$. By Lemma \ref{lem:address_geometry}(4) and Lemma \ref{lem:cross_subcube_edges}, each endpoint of $e_{q,k}$ can receive its additional incident edge only by restoring $e_{q,k}$ or by using the corresponding clause-variable edge. Since no occurrence is activated, $C_{\mathrm{occ}}$ contains none of the six clause-variable edges. It must therefore contain all three cut edges $e_{q,1},e_{q,2},e_{q,3}$, closing the three-cut clause forest $F_q$ into a separate cycle component and contradicting that $C_{\mathrm{occ}}$ is a single cycle.

Therefore, every clause has at least one activated occurrence. By the preceding argument, each such occurrence is a true literal under the assignment defined above. Hence every clause of $\Phi$ is satisfied, and $\Phi$ is satisfiable.
\end{proof}

\section{Completeness and complexity}

The preceding section has established that every Hamiltonian cycle containing $P(\Phi)$ induces a satisfying assignment of $\Phi$. We now prove the converse. Our task is therefore to select additional edges, according to a satisfying assignment $\sigma$ of $\Phi$, so that the resulting graph is a Hamiltonian cycle containing every edge of $P(\Phi)$. 

\begin{lemma}\label{lem:complete}
For every satisfying assignment of $\Phi$, there exists a Hamiltonian cycle of $Q_{a+d}$ containing every edge of $P(\Phi)$.
\end{lemma}

\begin{proof}
Fix a satisfying assignment $\sigma$ of $\Phi$, and put $\eta_i=\sigma(x_i)$ for $1\le i\le t$. First, we construct a Hamiltonian cycle $C_{\mathrm{occ}}$ covering all vertices in the occupied local subcubes and containing every edge of $G_{\mathrm{occ}}(\Phi)$.

For each variable $x_i$, let $\zeta = \zeta_{1-\eta_i}$ and $\zeta' = \zeta_{\eta_i}$ (so $\zeta = 46$ and $\zeta' = 54$ when $\eta_i=0$, and vice versa when $\eta_i=1$). Temporarily restore the deleted internal edges corresponding to $\{150,182\}$ in the bridge-incident terminal $A_8$-subcubes so that every terminal subcube at $v_{i,0}$ and $v_{i,L_i}$ contains the full two-path forest $A_8$ of Lemma \ref{lem:A1}. We assemble a cycle for variable $x_i$ as follows:
\begin{enumerate}
\item[(1)] Along all route links $0 \le j < L_i$, the prescribed common-channel edges join the common-channel paths of all local subcubes on the route into a single path running from $\Lambda(i,0,38)$ in $Q_d(v_{i,0})$ to $\Lambda(i,L_i-1,38)$ in $Q_d(v_{i,L_i})$ (since $\chi_0 = \chi_{L_i-1} = 14$ as $L_i$ is odd).
\item[(2)] On every route link $0 \le j < L_i$, select the Boolean candidate cross-subcube edge at port $\zeta$. Inside each internal $W_9$-subcube at $v_{i,p}$ ($0 < p < L_i$), select the internal residual edge $\{\Lambda(i,p-1,\zeta'), \Lambda(i,p,\zeta')\}$ (corresponding to $\{46,302\}$ or $\{54,310\}$ in Lemma \ref{lem:A2}). Together with the prescribed Boolean paths in the local subcubes, these edges form a second disjoint path running from $\Lambda(i,0,\zeta')$ in $Q_d(v_{i,0})$ to $\Lambda(i,L_i-1,\zeta')$ in $Q_d(v_{i,L_i})$.
\item[(3)] At the initial subcube $Q_d(v_{i,0})$ and the terminal subcube $Q_d(v_{i,L_i})$, close the two paths into a single cycle spanning the variable route of $x_i$ by adding the terminal residual edges $\{\Lambda(i,0,38), \Lambda(i,0,\zeta')\}$ and $\{\Lambda(i,L_i-1,38), \Lambda(i,L_i-1,\zeta')\}$ from $R(A_8) = \{\{14,46\},\{38,46\},\{38,54\}\}$.
\end{enumerate}

Next, for each $1 \le i < t$, remove the temporarily restored internal edges $\{\Lambda(i,L_i-1,150), \Lambda(i,L_i-1,182)\}$ in $Q_d(v_{i,L_i})$ and $\{\Lambda(i+1,0,150), \Lambda(i+1,0,182)\}$ in $Q_d(v_{i+1,0})$. Join the exposed endpoints across the two-cut bridge forest in $Q_d((b_i,0,00))$ via the four prescribed bridge-attachment edges. Since each bridge splices two distinct cycles by replacing one internal edge in each with two disjoint connecting paths, doing this for all $1 \le i < t$ merges the $t$ variable cycles into a single cycle covering all variable and bridge subcubes.

For each clause $C_q$, choose a literal occurrence $(q,k)$ that is true under $\sigma$, and put $i=i(q,k)$ and $p=p(q,k)$. Then $\eta_i=b(q,k)$, so the residual edge $\{\Lambda(i,p-1,\zeta_{b(q,k)}),\Lambda(i,p,\zeta_{b(q,k)})\}$ was selected in step~(2). Replace this edge by $g^-_{q,k}$, the Hamiltonian path $F_q+\{e_{q,k'}:k'\ne k\}$ through the clause subcube, and $g^+_{q,k}$. Each replacement incorporates a new clause subcube and preserves a single cycle. After all clauses have been incorporated, the resulting Hamiltonian cycle $C_{\mathrm{occ}}$ spans the occupied vertices and contains every edge of $G_{\mathrm{occ}}(\Phi)$.

Finally, apply to $C_{\mathrm{occ}}$ the exact sequence of path substitutions from Section~3.7 used to obtain $P(\Phi)$ from $G_{\mathrm{occ}}(\Phi)$, maintaining a current cycle throughout. At each step, the selected internal edge $xy$ of the current prescribed forest belongs to the current cycle. Replacing $xy$ by the two prescribed cross-subcube edges $xx',yy'$ and the prescribed Hamiltonian path through the newly incorporated subcube preserves a single cycle containing the updated prescribed forest. After all substitutions, the resulting graph is a Hamiltonian cycle of $Q_{a+d}$ containing $P(\Phi)$.
\end{proof}

To complete the complexity analysis, it remains to verify that the reduction can be carried out in polynomial time and that the decision problem itself belongs to $\mathrm{NP}$.

\begin{lemma}\label{lem:complexity}
The mapping $\Phi\mapsto(a+d,P(\Phi))$, where $P(\Phi)$ is given by its edge list, is computable in polynomial time. Moreover, the \textsc{spanning linear-forest hamiltonian-cycle extension} problem belongs to $\mathrm{NP}$.
\end{lemma}

\begin{proof}
Let $K=\lceil\log_2(H+1)\rceil$, where $H=3m+t\le6m$. Here $H\ge6$ and $K\ge3$. We have $r_b=K$, $h_c,h_v\le K$, and $h_\omega\le K+1$. Therefore $B\le5K+8$ and $a\le6K+10$. Each leg requires at most $B$ changes in the base-address block, at most $2r_b$ changes in the route-tag block for the transitions $0\to z$ and $z\to0$, and four changes in the lane label. Thus its length is at most $B+2r_b+4\le7K+12$. Consequently,
\[
L_{\max}\le H(7K+12)<2^{2K+3},
\]
using $H<2^K$ and $7K+12\le2^{K+3}$ for $K\ge3$. Hence $r=\lceil\log_2 L_{\max}\rceil\le2K+3$, $d=8+r+3h_v\le5K+11$, and $n\le11K+21$. In particular, $2^n\le2^{11K+21}<2^{32}(H+1)^{11}$. Since $P(\Phi)$ has maximum degree two, it contains at most $2^n$ edges. Each edge is specified by two $n$-bit endpoints, so its explicit edge list requires $O(n2^n)=O((H+1)^{11}\log(H+1))$ bits. As $H\le6m$, the output length is polynomial in the bit length of $\Phi$.

The variable routes and prescribed cross-subcube joins can be generated by the rules in Sections 3.1--3.3. The $A_8$- and $W_9$-subcubes given in Lemmas~\ref{lem:A1} and \ref{lem:A2} and their expansions can be constructed in $O(2^d d^{O(1)})$ time per subcube. Lemma~\ref{lem:small-matching-extension} constructs each clause and bridge cycle within the same time bound. The breadth-first forest can be generated by inspecting at most $a2^a$ address neighbors. The reserved edges can be selected and the path substitutions of Lemma \ref{lem:unused_subcubes} performed in $O(2^d n^{O(1)})$ time per address. The prescribed edge set can be maintained throughout the construction and then output as an explicit edge list. Since there are at most $2^a$ local subcubes, the construction takes $O(2^n n^{O(1)})$ time after reading and relabeling the formula. The bounds above therefore show that the total running time is polynomial in the bit length of $\Phi$.

For membership in $\mathrm{NP}$, consider any instance $(n,P)$ of the problem. Since every vertex has prescribed degree at least one, $2^n\le2|P|$. A certificate is a cyclic ordering of all $2^n$ vertices, requiring $O(n2^n)=O(n|P|)$ bits, which is polynomial in the input length. In polynomial time, we can verify that every vertex occurs exactly once, that consecutive vertices, including the last and first, are adjacent in $Q_n$, and that the cycle contains every edge of $P$. Hence, the problem belongs to $\mathrm{NP}$.
\end{proof}

We now combine Lemmas~\ref{lem:soundness}, \ref{lem:complete}, and~\ref{lem:complexity} to prove Theorem~\ref{thm:main}.

\begin{proof}[Proof of Theorem~\ref{thm:main}]
The 3-SAT problem in which every clause contains three literals on three distinct variables is NP-complete. Given a formula $\Phi$ in this problem, the construction in Section 3 produces an instance $(a+d,P(\Phi))$ of the \textsc{spanning linear-forest hamiltonian-cycle extension} problem, where $P(\Phi)$ is a spanning linear forest of $Q_{a+d}$ and every vertex has prescribed degree either one or two.

By Lemmas~\ref{lem:soundness} and~\ref{lem:complete}, $\Phi$ is satisfiable if and only if there exists a Hamiltonian cycle of $Q_{a+d}$ containing every edge of $P(\Phi)$. By Lemma~\ref{lem:complexity}, the mapping $\Phi\mapsto(a+d,P(\Phi))$ is computable in polynomial time. Hence, the \textsc{spanning linear-forest hamiltonian-cycle extension} problem is NP-hard. Moreover, by the same complexity lemma, it belongs to $\mathrm{NP}$. Therefore, the problem is NP-complete.
\end{proof}

We conclude this section by deriving two further
NP-hardness results for broader classes of prescribed edge sets. Both problems use the same input representation and Hamiltonian-cycle extension question as the \textsc{spanning linear-forest hamiltonian-cycle extension} problem, but impose weaker structural restrictions on the prescribed edge set.

\noindent\textsc{Linear-Forest Hamiltonian-Cycle Extension}

\smallskip
\noindent\textit{Instance:} An integer $n\ge 2$,
encoded in binary, and an explicitly listed edge
set $P\subseteq E(Q_n)$ such that $(V(P),P)$ is a
linear forest, where $V(P)$ denotes the set of
vertices incident with edges in $P$.

\smallskip
\noindent\textit{Question:} Does $Q_n$ contain a
Hamiltonian cycle $C$ such that $P\subseteq E(C)$?

\noindent\textsc{Prescribed-Edge Hamiltonian-Cycle Extension}

\smallskip
\noindent\textit{Instance:} An integer $n\ge 2$,
encoded in binary, and an explicitly listed edge
set $P\subseteq E(Q_n)$.

\smallskip
\noindent\textit{Question:} Does $Q_n$ contain a
Hamiltonian cycle $C$ such that $P\subseteq E(C)$?

\begin{corollary}\label{cor:broader}
The \textsc{linear-forest hamiltonian-cycle extension} problem is NP-hard. Moreover, the
\textsc{prescribed-edge hamiltonian-cycle extension} problem is also NP-hard.
\end{corollary}

\begin{proof}
The reduction constructed in the proof of
Theorem~\ref{thm:main} maps every source 3-SAT instance $\Phi$ to an instance $(a+d,P(\Phi))$ in which $P(\Phi)$ is a spanning linear forest and every vertex has prescribed degree one or two. Consequently, every instance produced by this
reduction satisfies the input restrictions of both
the \textsc{linear-forest hamiltonian-cycle extension} problem and the unrestricted
\textsc{prescribed-edge hamiltonian-cycle extension} problem.

By Lemmas~\ref{lem:soundness} and~\ref{lem:complete}, $\Phi$ is satisfiable if and only if $P(\Phi)$ is contained in a Hamiltonian cycle of $Q_{a+d}$. Moreover, the reduction is computable in polynomial time by
Lemma~\ref{lem:complexity}. Hence, the same reduction establishes NP-hardness for both problems.
\end{proof}

The same reduction establishes NP-hardness for both problems when the host hypercube is given by explicit vertex and edge lists, since the constructed dimension satisfies $n=O(\log|\Phi|)$. Under this representation, a Hamiltonian cycle is a polynomial-size certificate that can be verified in polynomial time, so both problems are NP-complete.
\section{Concluding Remarks}

We prove that the Hamiltonian-cycle extension problem on hypercubes is NP-complete when the prescribed edge set is a spanning linear forest in which every vertex has prescribed degree one or two. The same reduction also establishes NP-hardness when the prescribed edge set is a linear forest not necessarily spanning, and even when no structural restrictions are imposed on the prescribed edge set. For sparse prescribed-edge instances in these more general settings, however, an explicit Hamiltonian cycle may have superpolynomial size relative to the input. Consequently, the usual explicit-cycle certificate does not establish membership in $\mathrm{NP}$ for these more general problems, although this does not rule out the existence of other polynomial-size certificates.

An especially restrictive special case of the prescribed-edge extension problem arises when $P$ is a matching. This is the setting of the Ruskey-Savage conjecture, which asks whether every matching of $Q_n$, for every $n\ge 2$, can be extended to a Hamiltonian cycle. Our reduction does not address this case, since the prescribed forests produced by our construction contain vertices of prescribed degree two and hence are not matchings. Investigating the Hamiltonian-cycle extension problem when the prescribed edge set is restricted to a matching is a natural direction for further research.

\bibliographystyle{elsarticle-num-alpha} 
\bibliography{ref}
\end{document}